# Decoherence of Quantum Emitters in hexagonal Boron Nitride

Jake Horder[1,2], Dominic Scognamiglio[1], Nathan Coste[1], Angus Gale[1], Kenji Watanabe[3], Takashi Taniguchi[4], Mehran Kianinia[1,2], Milos Toth[1,2] and Igor Aharonovich[1,2]

1. School of Mathematical and Physical Sciences, University of Technology Sydney, Ultimo, New South Wales 2007, Australia
2. ARC Centre of Excellence for Transformative Meta-Optical Systems, University of Technology Sydney, Ultimo, New South Wales 2007, Australia
3. Research Center for Electronic and Optical Materials, National Institute for Materials Science, 1-1 Namiki, Tsukuba 305-0044, Japan
4. Research Center for Materials Nanoarchitectonics, National Institute for Materials Science, 1-1 Namiki, Tsukuba 305-0044, Japan305-0044, Japan
milos.toth@uts.edu.au; igor.aharonovich@uts.edu.au

**Abstract**
Coherent quantum emitters are a central resource for advanced quantum technologies. Hexagonal boron nitride (hBN) hosts a range of quantum emitters that can be engineered using techniques such as high-temperature annealing, optical doping, and irradiation with electrons or ions. Here, we demonstrate that such processes can degrade the coherence—and hence the functionality—of quantum emitters in hBN. Specifically, we show that hBN annealing and doping methods that are used routinely in hBN nanofabrication protocols give rise to decoherence of B-center quantum emitters. The decoherence is characterized in detail, and attributed to defects that act as charge traps which fluctuate electrostatically during SPE excitation and induce spectral diffusion. The decoherence is minimal when the emitters are engineered by electron beam irradiation of as-grown, pristine flakes of hBN, where B-center linewidths approach the lifetime limit needed for quantum applications involving interference and entanglement. Our work highlights the critical importance of crystal lattice quality to achieving coherent quantum emitters in hBN, despite the common perception that the hBN lattice and hBN SPEs are highly-stable and resilient against chemical and thermal degradation. It underscores the need for nanofabrication techniques that are minimally invasive and avoid crystal damage when engineering hBN SPEs and devices for quantum-coherent technologies.

**Main text**
Advanced quantum technologies are at the forefront of modern research, with applications such as quantum computing [1] and long-range quantum networks [2,3] garnering significant interest in the wider scientific and commercial communities. A key resource for these technologies are coherent single photon emitters (SPEs) [4,5]. Semiconductor quantum dots and color centers in diamond have long been favored as solid-state SPE platforms [6]. However, the material hosts of these emitters which make them suitable for scalable on-chip technologies are also a potential source of environmental noise in the form of phonon-induced dephasing and spectral diffusion, which degrade the coherence of quantum emitters.

Recently, hexagonal boron nitride (hBN) has gained interest as a van der Waals material host of SPEs with high brightness, photon purity and photostability [7,8]. Remarkably, the linewidths of hBN SPEs have been shown to approach the Fourier-transform limit (FTL) under resonant excitation [11,12] without the need for active decoherence mitigation strategies such as optical repumping, which are typically needed to realize long coherence times in SPE hosts such as diamond [11–14]. However, despite their compelling photophysical characteristics, scalable integration of hBN SPEs with nanostructures and devices has proved challenging [15,16] because some properties, such as emission wavelength and SPE spatial location, are difficult to control/engineer deterministically and consistently. A notable exception is the B-center SPE, which can be engineered on-demand using a site-specific electron beam irradiation technique, and exhibits a highly consistent zero-phonon line (ZPL) at 436 nm [17–19]. These properties have enabled rapid, timely demonstrations of B-center integration in photonic nanostructures [20,21] and measurements of B-center linewidths under resonant excitation [22].

However, most approaches to engineering hBN quantum emitters and integrating them in devices involve invasive material processing and nanofabrication techniques [23–25]. Such techniques are likely to generate lattice defects that may cause decoherence and compromise the utility of SPEs. This is problematic because, despite the decoherence, the emitters can, in principle, retain basic photophysical characteristics which are often used as metrics of "quality" and suitability for quantum technologies. Such metrics include high brightness and emission stability under non-resonant excitation, as well as photon purity and polarization visibility [26,27].

Here we show that two common hBN processing methods—annealing and carbon doping—give rise to decoherence of B-center quantum emitters which limits their suitability for applications based on interference and entanglement. Conversely, the decoherence is minimal when B-centers are engineered by electron beam irradiation of as-grown, pristine flakes of hBN. Decoherence is evaluated by emission linewidths, measurements of power broadening and emission stability under resonant excitation. It is attributed to lattice defects that act as charge traps which fluctuate electrostatically during SPE excitation.

Our results demonstrate that the coherence of B-centers—the most promising hBN SPEs investigated to date—can be compromised by common hBN processing methods. Indeed, the doping and annealing methods are used routinely to engineer hBN flakes and devices that contain not only B-centers [21], but also emitters with ZPLs in the UV [28,29] and visible [27,30] spectral ranges. Broadly, our work highlights the need for minimally-invasive hBN processing and nanofabrication protocols for engineering of coherent quantum emitters and devices. One such protocol—electron beam irradiation—is demonstrated here to yield B-center SPEs with stable, narrow linewidths observed in the absence of active decoherence mitigation strategies.

Three types of mechanically-exfoliated flakes of hBN were investigated:
i) *Pristine* (Figure 1a): hBN flakes exfoliated from crystals grown under high-pressure high-temperature (HPHT) conditions at the National Institute for Materials Science (NIMS). These flakes are representative of samples used broadly in the hBN literature.
ii) *Annealed* (Figure 1b): pristine flakes that were annealed in air at 850 ºC. This annealing treatment is representative of those used routinely to fabricate various types of SPEs in hBN, and to increase the fidelity of an electron beam technique used to engineer B-centers deterministically [25,31–33].
iii) *C-doped* (Figure 1c): pristine flakes that were annealed at 2100 ºC in a carbon furnace under flowing $N_2$ gas. Carbon doping is sometimes used to further enhance the fidelity of the above B-center electron beam engineering technique [23,34]. Carbon dopants are also used to generate UV emitters in hBN [35], and they play a role in SPEs that emit in the visible spectral range [36,37].

B-centers were engineered in the above three hBN types by the electron beam irradiation technique detailed in the methods section. For context, we note that pristine hBN is considered unappealing for B-center engineering by electron beam irradiation because the process fidelity (i.e., success rate) is relatively low [31]. Conversely, the fidelity is known to increase when using annealed and C-doped flakes of hBN. This is illustrated in Figure S1 by representative photoluminescence (PL) maps of the three types of hBN, taken after electron beam irradiation. Relatively few B-centers are seen on the pristine flake of hBN, whereas the other two samples each host numerous emitters. The enhanced fidelity is desirable but it comes at the cost of reduced B-center coherence, demonstrated by the spectra in Figure 1d-f. The spectra are representative resonant photoluminescence excitation (PLE) scans from a typical B-center in each sample type. The Lorentzian full-width-at-half-maximum (FWHM) linewidths—which are inversely proportional to the coherence times—are 0.88 GHz for pristine hBN, 1.15 GHz for annealed hBN, and 14.92 GHz for C-doped hBN. These values demonstrate an inverse relationship between the suitability of hBN flakes for high-fidelity B-center engineering by electron beam irradiation and the coherence of the resulting B-center quantum emitters. Rephrasing, the hBN flakes that are most favorable for SPE engineering and device nanofabrication are least suitable for applications that require highly-coherent SPEs.

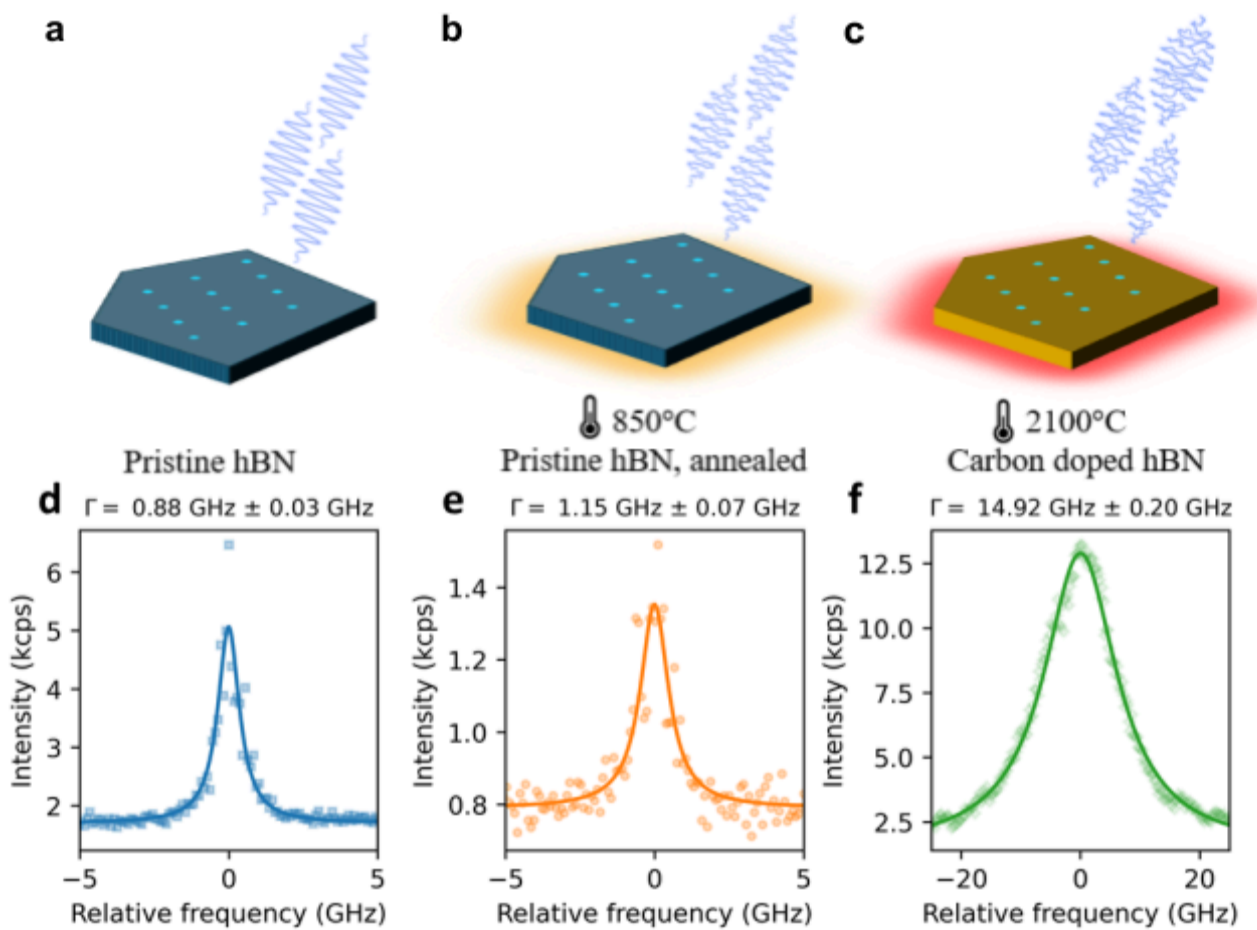

*Figure 1. Linewidths of B-centers in three types of hBN. a-c) B-centers were characterized in pristine hBN (a), hBN that was annealed in air at 850 ºC (b), and hBN that was doped with carbon. d-e) Average spectra obtained*

*from several PLE scans across the ZPL of a typical B-center in pristine hBN (d), and annealed hBN (e). In each case, the excitation power is 2 µW and the data are fit with a Lorentzian function to extract the FWHM linewidths shown in the figure. f) Average spectrum obtained from several PLE scans across the ZPL of a typical B-center in C-doped hBN, using an excitation power of 10 µW.*

Next, we characterize in detail the B-centers in the pristine and annealed hBN. Emitters in the C-doped hBN were all found to be approximately two orders of magnitude broader in linewidth than the Fourier transform (FT) limit of ~80 MHz. This is inadequate for applications underpinned by coherence, and we therefore discount the C-doped flakes from further analysis.

Figure 2a-b compares directly the resonant power-dependent properties of the brightest B-center found in pristine and annealed hBN. The saturated emission intensity $I_\infty$, obtained by fitting the data in Figure 2a using the saturation function $I(P) = I_\infty P / (P + P_{sat})$, is 49 kcps and 18 kcps for the pristine and annealed flakes, respectively. This difference is important because a reduced emission rate is indicative of spectral diffusion (discussed in detail below) and because high photon rates are critical for advanced applications, such as those involving two-photon interference.

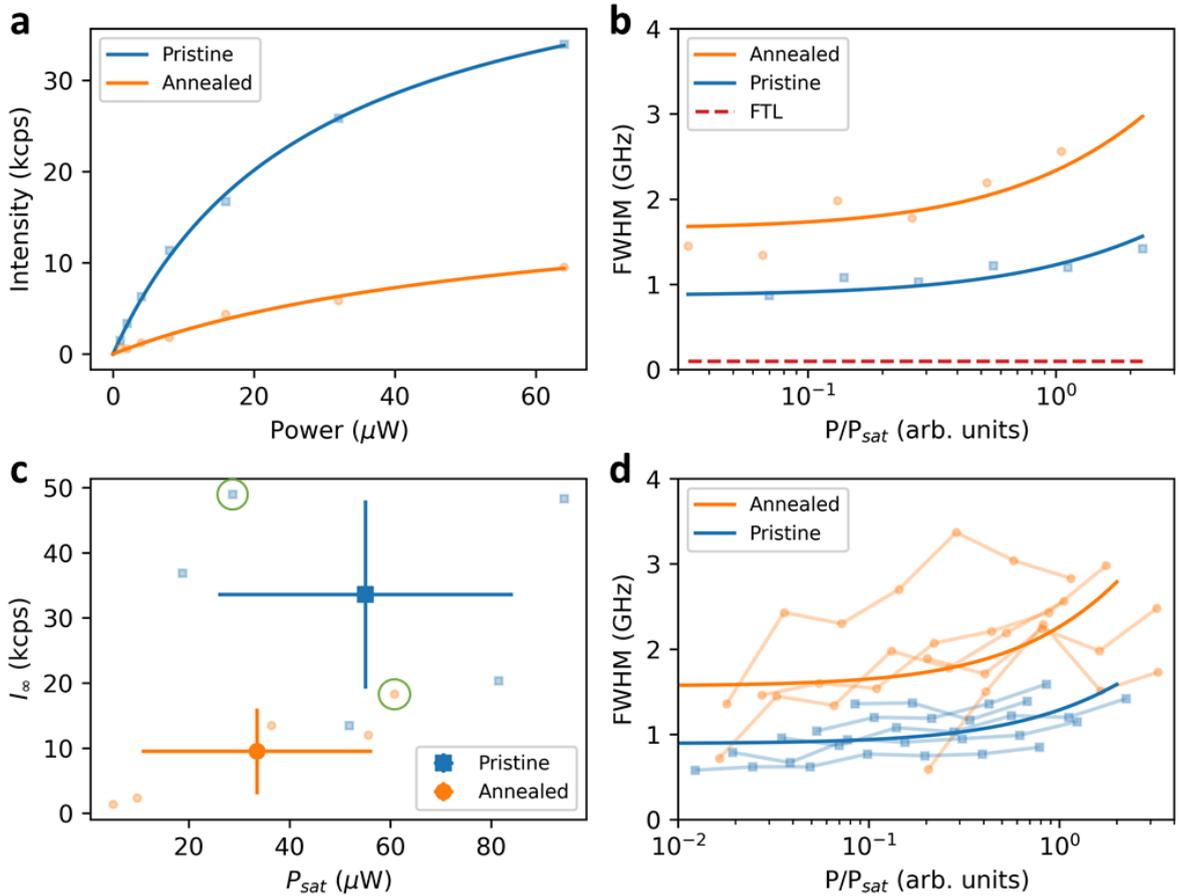

*Figure 2. Emission intensity and linewidths of B-centers in pristine hBN (blue) and annealed hBN (orange), as a function of excitation power. a) Power saturation behavior of the brightest B-centers found in pristine and annealed hBN (circled in green in (c)). The saturation power $P_{sat}$ is 28 µW and 61 µW, and the corresponding saturated intensities are 49 kcps and 18 kcps for the pristine for annealed cases, respectively. b) FWHM of the two B-centers in (a). The red dashed line is the Fourier transform limited (FTL) linewidth. c) Saturation power versus saturated intensity for five B-centers in each sample type. On average, B-centers in pristine hBN are brighter than in annealed hBN. The B-centers in pristine hBN appear to have slightly higher saturation power, although the difference in average values for $P_{sat}$ is within the standard deviation. d) Average power-dependent*

FWHM of all B-centers measured in both types of hBN. On average, the B-center linewidths are approximately power-independent at low powers, and both broaden at high excitation powers. The broadening is greater for B-centers in annealed hBN.

Figure 2b shows plots of linewidth versus excitation laser power, fitted using the power broadening function $\Gamma(P) = \Gamma_0(1 + P/P_{sat})^{1/2}$. The fits indicate that the emission from the annealed hBN is consistently twice as broad at all powers employed in the experiments. Extrapolating to the limit of very low power, we find that both emitters have a power-independent linewidth $\Gamma_0$ that is approximately one order of magnitude greater than the FTL linewidth, shown as a dashed red line in Figure 2b. This suggests that, in the limit of low excitation power, the Lorentzian linewidths are, in both cases, dominated by decoherence caused by phonons and/or fast spectral diffusion, as is discussed further below.

In Figure 2c-d we compare the power-dependent intensity and linewidth data from five B-centers in pristine hBN and five in annealed hBN. The average saturation power $P_{sat}$ and average saturated intensity $I_\infty$ are shown in Figure 2c, where error bars represent one standard deviation. The emitters in pristine hBN are significantly brighter, on average, indicating more time spent on-resonance with the laser – i.e., this indicates that spectral diffusion of B-centers in annealed hBN is greater, resulting in less time spent, on average, on-resonance with the excitation laser.

At low powers, the average B-center linewidths are approximately independent of the excitation power in both the pristine and annealed flakes, and both broaden at high excitation powers (Figure 2d). The broadening is greater for B-centers in annealed hBN. A reduced coherence correlating with laser intensity has been attributed previously to laser-induced local heating [38], suggesting that the annealed flake contains more defects that facilitate heating via laser absorption.

We note that the high degree of noise/scatter seen in Figure 2c,d is expected because, as we show below, it is caused by spectral diffusion, which is a local effect. The magnitude of spectral diffusion experienced by any one SPE is determined by the local crystal environment – specifically, it is determined by the local densities of defects which vary throughout each flake of hBN.

Next, we investigate in detail the spectral diffusion of B-centers in pristine and annealed hBN under resonant excitation. Spectral diffusion likely plays a substantial role in the linewidth broadening observed in our data since phonon dephasing has been shown previously to broaden B-center linewidths to only ~200 MHz (using the spectral hole burning technique) [39]. Conversely, the linewidths measured here in pristine and annealed hBN are several hundred MHz broader than this, and several tens of GHz in the case of the C-doped hBN.

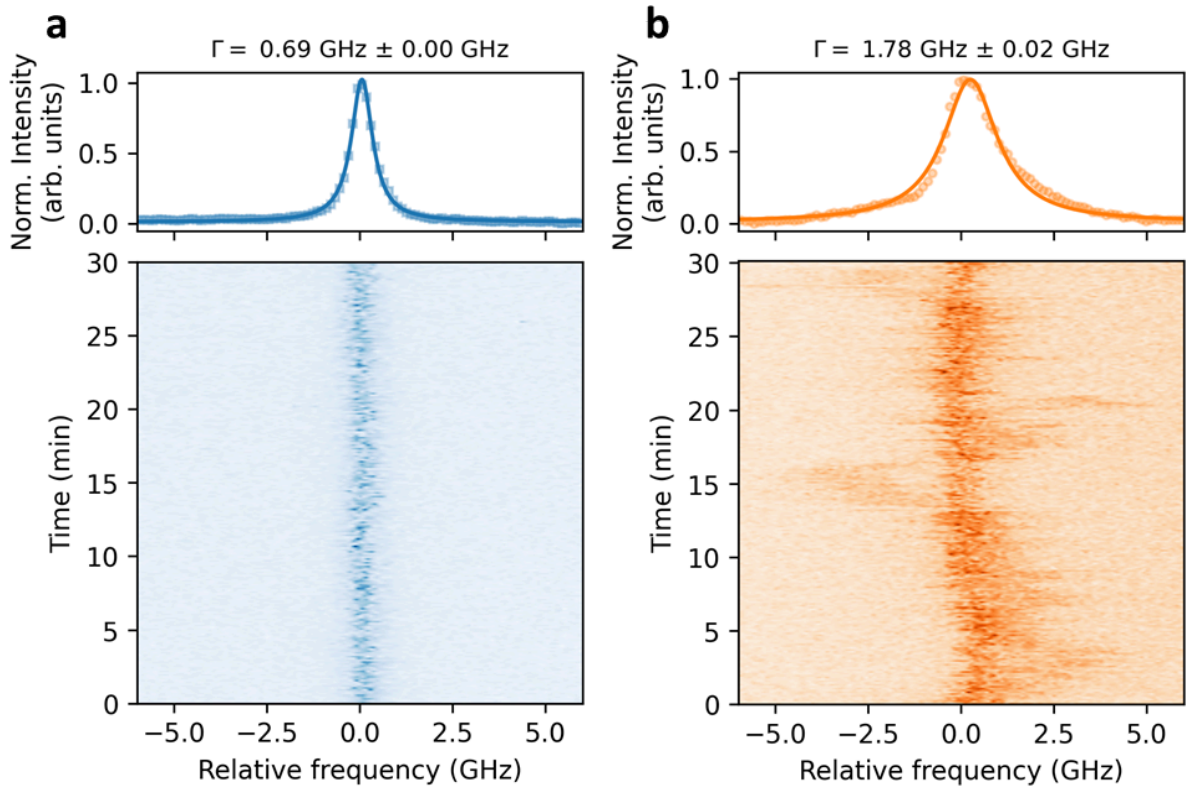

*Figure 3. Spectral stability of B-centers in pristine and annealed hBN. a) Average PLE intensity over 30 minutes with a Lorentzian fit (top panel) and constituent PLE scans resolved in time (bottom panel) for a B-center in pristine hBN. The resonant laser power is 10 μW. b) The same measurement for a B-center in annealed hBN.*

We note that the lineshapes observed in this study are appropriately Lorentzian rather than Gaussian (see Figure S2), indicating that the broadening is likely caused by rapid spectral diffusion – i.e. fluctuations of the ZPL energy at a rate that is much greater than the spontaneous emission rate. Such fluctuations can be caused by electric fields generated by charge traps in the vicinity of each B-center, which fluctuate electrostatically during laser excitation. This implies that the observed differences in linewidth broadening are likely caused by corresponding differences in the densities of defects introduced into the hBN lattice during annealing and C-doping.

In Figure 3a, a highly stable resonance is observed for a B-center in pristine hBN, where the Lorentzian linewidth from PLE scans averaged over 30 minutes is only 690 MHz, equivalent to the single scan average (see Figure S3). In contrast, the emitter in the annealed sample shown in Figure 3b exhibits a slightly broader scan-by-scan linewidth in addition to wandering of the ZPL at rates spanning a few minutes to tens of minutes. This longer term spectral diffusion manifests in a slight deviation from the Lorentzian shape of PLE lineshape (seen as a discrepancy between the fit shown in the top panel of Figure 3b and the experimental data). The long term spectral stability is important for advanced quantum optics experiments and applications in which the resonant laser is not scanned but instead fixed at the central wavelength of the ZPL. In this configuration, the presence of slow spectral diffusion seen in Figure 3b would manifest as blinking of the SPE—i.e., an emission instability due to changes in brightness at time scales spanning seconds or minutes. This fixed excitation condition is characterized directly in Figure 4 for B-centers in pristine and annealed hBN. PLE scans at saturation power were first used to determine the central wavelength of the brightest B-center found in pristine hBN. The resonant laser was then fixed at this wavelength and intensity time traces were recorded over two minutes for a range of powers. This was repeated for the brightest observed B-center in annealed hBN.

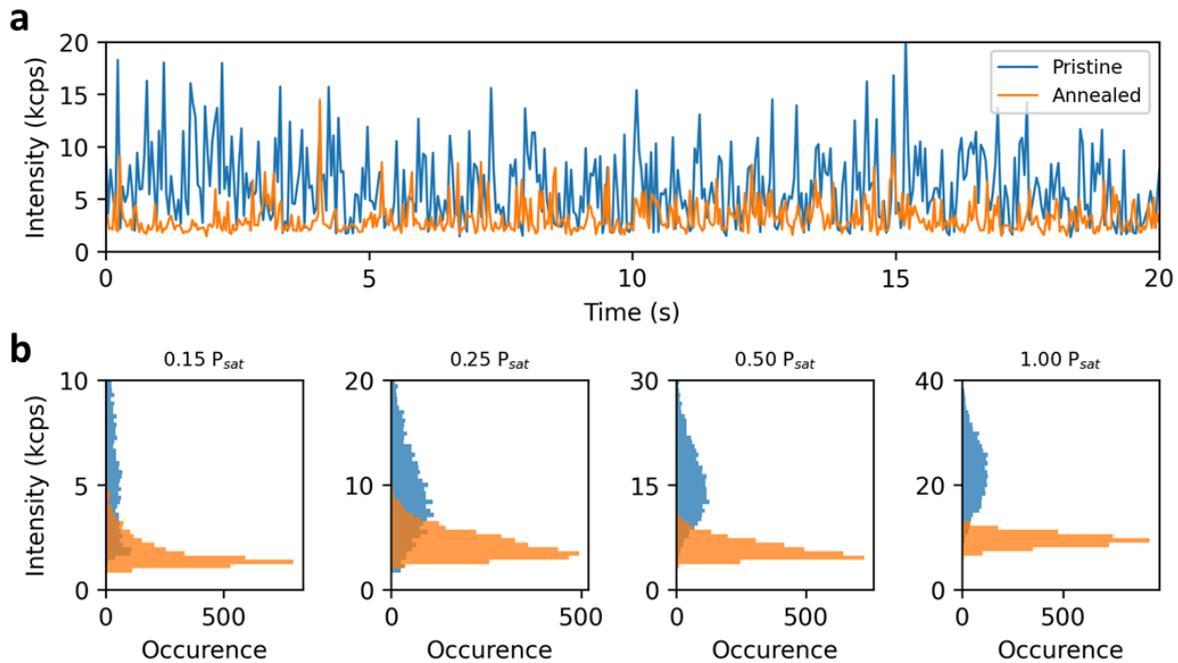

*Figure 4. Intensity stability B-centers in pristine and annealed hBN under fixed resonant excitation. a) Time-resolved intensity from a B-center in pristine hBN and a B-center in annealed hBN, using 15% of the saturation power. The emitter in pristine hBN has greater average intensity and is more consistently bright. b) Intensity histograms from 2 minutes of fluorescence at various fractions of saturation power, from the same emitter in pristine hBN (blue) and in annealed hBN (orange) as in (a). Above $0.15P_{sat}$, the pristine case has a normally-distributed intensity, while the annealed case is skewed toward a dim intensity.*

Figure 4a shows a snapshot of the time-dependent intensity under an excitation power of $0.15P_{sat}$. While there is considerable variation in intensity for the pristine case, the count rate is mostly far greater than the detector dark count rate. For the annealed case, there are relatively long periods of no emission, and this is reflected in the skewed intensity histogram shown in Figure 4b. The skewed distribution persists across all powers up to $P_{sat}$ for the B-center in annealed hBN. The prolonged dim periods are likely due to spectral jumps of the emitter out of resonance with the laser (since the B-center is not believed to feature metastable states [40]). A stronger quantitative determination of various rates of spectral diffusion could be explored through photon correlations [41–43]. These are, however, beyond the scope of the present study (and would not be insightful broadly since they would characterize, in detail, the local environments of the specific two SPEs presented in Figure 4).

Summarizing, we have demonstrated that common processing steps used in hBN SPE and device engineering protocols can reduce emitter coherence. The employed carbon-doping protocol leads to linewidth broadening that renders B-center quantum emitters unsuitable for quantum-coherent technologies. Annealing in air at 850 ºC leads to moderate linewidth broadening, and spectral diffusion that occurs over a broad range of timescales. Spectral diffusion that is fast relative to the excited state lifetime is likely responsible for the observed Lorentzian broadening of PLE lines measured under resonant excitation. Spectral diffusion at medium and long timescales manifests as non-Lorentzian broadening and peak shifts/jumps in PLE spectra, and also as intensity fluctuations under fixed-wavelength resonant excitation. The resulting decoherence observed in annealed and carbon-doped hBN is consistent with spectral diffusion, which is likely caused by defects (i.e., charge traps) introduced into the hBN lattice during the annealing and doping procedures.

In C-doped hBN, the observed spectral diffusion leads to significant linewidth broadening (Figure 1f) and hence severe decoherence. However, despite this, the B-center SPEs retain some useful properties, including high brightness under non-resonant excitation, high polarization contrast, and high photon purity. This makes them suitable for several applications, including some forms of quantum key distribution [44] and superresolution microscopy [45]. However, the observed decoherence renders them unsuitable for applications that involve interference and entanglement.

Conversely, B-centers in pristine hBN exhibit long coherence times, indicated by near-FTL PLE linewidths (Figure 1d, 3a), even though an electron beam irradiation treatment was used to engineer the investigated B-center SPEs. This indicates that the electron beam technique can be "minimally-invasive" with regards to SPE coherence times. It can yield emitters that are promising for advanced quantum optics applications such as photonic quantum computation [46] and boson sampling [47].

The investigated annealed hBN samples are appealing in that the annealing treatment increases the fidelity of the electron beam irradiation process (Figure S1). It, however, degrades B-center coherence and brightness under resonant excitation, but the observed PLE linewidths are similar to those of B-centers in pristine hBN (Figure 2b). Further work to optimize the annealing conditions—temperature, duration and gaseous environment—may yield recipes that minimize crystal damage, or perhaps even improve crystal quality via thermally-induced defect annihilation.

In conclusion, we have demonstrated that two routine hBN processing techniques—annealing in air at ~850 °C and heavy carbon-doping—degrade the coherence of B-center quantum emitters. Our results illustrate that, despite the common perception of hBN as a highly-inert material that is resilient against chemical and thermal degradation, common processing steps can render hBN SPEs unsuitable for quantum-coherent applications. Our findings highlight the need for hBN processing and nanofabrication techniques that are minimally invasive and avoid crystal damage. One such example is electron-beam-irradiation of pristine hBN, which can be used to engineer coherent B-center SPEs with near-FTL linewidths.

## Methods
**Sample preparation:**
Pristine: Bulk hBN grown using the HPHT method was used to exfoliate a thin hBN flake onto a $SiO_2$/Si substrate. Cathodoluminescence spectrometry was then performed to identify suitable hBN flakes for the creation of B-centers.
Annealed: Pristine flakes were treated to a low ramping annealing recipe up to 850 °C over 4 hours.
C-doped: Carbon doped bulk hBN grown using the HPHT (see Reference [35]) was used to exfoliate a thin hBN flake onto a $SiO_2$/Si substrate.

**Electron beam irradiation:**
Electron beam irradiations were undertaken using a Thermo Fisher Scientific Helios G4 Dual Beam microscope. A focused electron beam (5 keV, 3.2 nA) was used to irradiate the hBN flakes. An 8 x 8 array with 3 μm spacing was patterned using the inbuilt software. All samples were irradiated in one sitting to ensure identical electron beam parameters. Irradiated flakes were then randomly selected from the pristine and C-doped samples, which were then transferred using the PVA stamp method to the annealed substrate, due to the difficulty of transferring hBN flakes post annealing. The substrate holding the three samples was then cleaned in a UV ozone bath (ProCleaner Plus, Bioforce Nanosciences Inc.) for 30 minutes.

**Photoluminescence experiments:**

The substrate containing all three samples was cooled to 5 K using a closed-loop cryostat with a window leading to a home-built confocal microscope setup. For all PLE experiments, resonant excitation was performed with a frequency doubled Ti:sapphire scanning laser (M Squared) and photons were collected from the phonon sideband (PSB) using a 442 nm long pass filter. Data was collected over 25 ms integration time. Scanning was performed at a rate of 1 GHz/s unless otherwise specified. Confocal scanning PL maps were obtained under 1 mW of 405 nm laser excitation.

**Acknowledgements**

We acknowledge financial support from the Australian Research Council (CE200100010, FT220100053, DP240103127), and the UTS node of the ANFF for access to nanofabrication facilities. K.W. and T.T. acknowledge support from the JSPS KAKENHI (Grant Numbers 21H05233 and 23H02052) and World Premier International Research Center Initiative (WPI), MEXT, Japan.